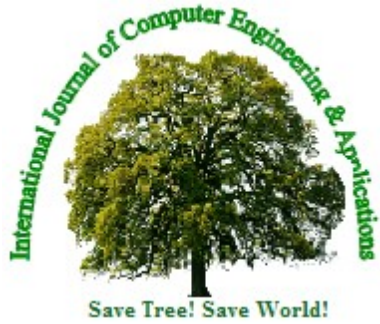

# ADDRESSING SECURITY CHALLENGES IN CLOUD COMPUTING

Abu Salim[1], Dr. Rajesh Kumar Tiwari[2], Dr. Sachin Tripathi[3]

[1] Department of Computer Science, GLNAIT, Mathura.
[2] Department of Computer Science, RVS College of Eng. & Tech, Jamshedpur.
[3] Department of Computer Science and Engineering, ISM Dhanbad.

**ABSTRACT:**

Cloud computing is a new computing paradigm which allows sharing of resources on remote server such as hardware, network, storage using internet and provides the way through which application, computing power, computing infrastructure can be delivered to the user as a service. Cloud computing unique attribute promise cost effective Information Technology Solution (IT Solution) to the user. All computing needs are provided by the Cloud Service Provider (CSP) and they can be increased or decreased dynamically as required by the user. As data and Application are located at the server and may be beyond geographical boundary, this leads a number of concern from the user prospective. The objective of this paper is to explore the key issues of cloud computing which is delaying its adoption.

# [I] NTRODUCTION

The researchers in academic and industry are doing intensive research to come back with better technology and as a result they have come up with cloud computing.

The grid and cloud computing are forms of distributed computing which are based on service provisioning model. In which user pay service provider for using their services.

A distributed computer system consists of multiple software components that are on multiple computers, but run as a single system. It can be considered as conventional networks of independent computers which solve a large problem by giving small parts of the problem to many computers and then combining the solutions.

With grid computing, you can provision computing resources as a utility that can be turned on or off. Cloud computing goes one step further with on-demand resource provisioning. It o removes the need to over-provision in order to meet the demands of millions of users.

Cloud computing provides computing as a service rather than a product, shared resources such as hardware and software are provided to computers and other devices (tablet, mobiles etc) as a utility over a network.

It takes advantage of high availability, high performance, scalable computing services that are provided on a self-service, pay-as-you-go basis over the Internet. A wide range of services are available today ranging from basic infrastructure hosting (IaaS), to full development platform hosting (PaaS) to applications hosting (SaaS).

Cloud Computing can improve time-to-market as a new server can be created or brought online quickly. It Helps users avoid costly up-front capital investments in the infrastructure. Increase the flexibility of both the business and the IT organization as allows firms to pay for excess capacity only when they need it.

As with many great opportunities, Cloud Computing also presents many challenges and risks which IT staff and decision-makers need to be aware of.

The Paper is organized as follows. Section II discuss the Key Characteristics of cloud computing. Section III discuss the services provided by cloud computing. Section IV discusses the service delivery model. Section V describe cloud provider and entities .Section VI discuss key security issues in cloud computing environments. Section VII discuss some cloud computing limitation, we conclude the paper in Section VIII.

## [II] CLOUD COMPUTING CHERISTRISTICS

Cloud computing poses five key characteristics, Utilizes three delivery model and four deployments Model.

**Key characteristics:-**

**2.1 On Demand Self-Service:** Customer can access /control computing resources automatically as needed.

**2.2 Ubiquitous Network Access:** User can access application/data with the help of different type of devices like Computer, Mobile, and Tablet.

**2.3 Resource Pooling:** Cloud Service Provider (CSP) can share resources (hardware/Software) to the different users. User can acquire or release resources as needed.

**2.4 Rapid Elasticity:** User can acquire or release resources quickly and automatically as needed.

**2.5 Measured Service:** Uses of the resources on the cloud can be monitored and the user is charged on the basis of the resources used.

# [III] CLOUD COMPUTING SERVICE DELIVERY MODELS

Cloud computing utilizes three delivery models by which different types of services are delivered to the end user. The three delivery models are the SaaS, PaaS and IaaS which provide infrastructure resources, application platform and software as services to the consumer. These service models also place a different level of security requirement in the cloud environment.

**3.1 Software as a Service (SaaS):** It is software deployment model in which application are deployed remotely on the server by the Cloud service provider and those services can be used by the cloud users. SaaS offer a number of benefits to the cloud users such as decrease upfront investment and operational cost. However, most of the organizations are still uncomfortable because of the number of concern like security, privacy and trust.

Examples of PaaS offering company are Salesforce, Gmail. Salesforce put cloud computing on the map with its cloud-based CRM SaaS solution. Gmail is another popular SaaS solution.

**3.2 Infrastructure as a Service (IaaS):** IaaS completely changes the way developers deploy their applications. Instead of spending big with their own data centers or managed hosting companies or co-location services and then hiring operations staff to get it going, they can just go to one of the IaaS providers, get a virtual server running in minutes and pay only for the resources they use. In short, IaaS and other associated services have enabled start-ups and other businesses focus on their core competencies without worrying about infrastructure of IT Services. IaaS provides only basic security; applications moving into the cloud will need higher levels of security and regulatory requirements.

Examples of PaaS offerings include force.com, Google App Engine or Microsoft Azure.

**3.3 Platform as a Service (PaaS):** PaaS is one layer above IaaS on the stack and abstracts away everything up to OS, middleware, etc. This offers an integrated set of developer environment that a developer can tap to build their applications without having any clue about what is going on underneath the service. It offers developers a service that provides a complete software development life cycle management, from planning to design to building applications to deployment to testing to maintenance. Everything else is abstracted away from the ''view'' of the developers. The dark side of PaaS is that, these advantages itself can be helpful for a hacker to leverage the PaaS cloud infrastructure for malware command and control and go behind IaaS applications.

Examples of IaaS offerings include Amazon Web Services, GoGrid, Rackspace.

## [IV] CLOUD COMPUTING SERVICE DEPLOYMENT MODELS

The Cloud Computing Utilizes Four Deployment Models.

**4.1 Private Cloud:** In this type of cloud computing, cloud infrastructure is used by a single organization which may be located on or off premises.

**4.2 Community Cloud:** In this type of cloud computing, cloud infrastructure are used by many organizations that have shared concerns (e.g., mission, security requirements, policy, and compliance considerations). It may be managed by the organizations or a third party and may exist on premise or off premise.

**4.3 Public Cloud:** In this type of cloud computing, cloud infrastructure is made available to the general public or a large industry group and is owned by an organization selling cloud services known as a Cloud Service Provider.

**4.4 Hybrid Cloud:** In this type of cloud computing, cloud infrastructure is a composition of two or more clouds (private, community, or public) that remain unique entities but are bound together by standardized or proprietary technology that enables data and application portability (e.g., cloud bursting for load-balancing between clouds).

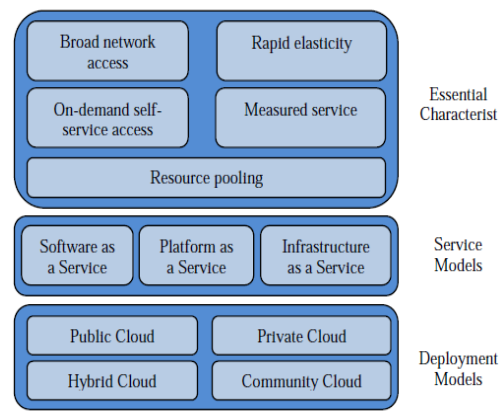

Fig. 1: NIST visual model for cloud computing
http://www.csrc.nist.gov/groups/SNS/cloud-computing/index.html

## [V] CLOUD COMPUTING USERS AND PROVIDERS

A number entities are involved in the Cloud computing like cloud user, cloud provider, cloud reseller and cloud service broker.

**5.1 Cloud Providers:** Includes Internet service providers, telecommunications companies, and large business process outsourcers that provide either the media (Internet connections) or infrastructure (hosted data centers) that enable consumers to access cloud services. Service providers may also include systems integrators that build and support data centers hosting private clouds and they offer different services (e.g., SaaS, PaaS, IaaS, and etc.) to the consumers, the service brokers or resellers.

**5.2 Cloud Service Brokers:** Includes technology consultants, business professional service organizations, registered brokers and agents, and influencers that help guide consumers in the selection of cloud computing solutions. Service brokers concentrate on the negotiation of the relationships between consumers and providers without owning or managing the whole Cloud infrastructure. Moreover, they add extra services on top of a Cloud provider's infrastructure to make up the user's Cloud environment.

**5.3 Cloud Resellers:** These are local company chosen by the cloud service provider to extend their business. They are act as a cloud provider in a region but actually they are providing services with the help of other Service Provider.

**5.4 Cloud Users:** They use the services offered by the cloud service provider, they are at the last level in this chain.

## [VI] KEY CHALLENGES IN CLOUD COMPUTING

This section will discuss key challenges faced by the cloud computing.

**6.1 Data Security:** In cloud computing security of data stored at the Cloud Service Provider (CSP) is a major concern. In traditional application data are stored on the computer which is reside at the customer premises. The physical and logical security should be provided by the customer. In the cloud computing environment the data are stored on shared environment which is not under the control of user. SaaS vendor must provide security to the data. They must use strong encryption technique fine grain access control to the data. The customer should have control over data life cycle, once the data is deleted by the user it should not be regained by SaaS provider by using any means. Data can be encrypted automatically before storage. Encryption provided by hard disk manufacturer is having less performance overhead.

The following test can be used to check security of data stored at the cloud service provider [1].

- Cross-site scripting[XSS]
- Access control weaknesses
- OS and SQL injection flaws
- Cross-site request forgery [CSRF]
- Cookie manipulation
- Hidden field manipulation
- Insecure storage

**6.2 Data Privacy:** Privacy is the ability of an individual or group to seclude themselves or information about themselves and thereby revels themselves selectively [2].

The data privacy is also a matter of concern in the cloud computing. Companies need to setup a group of employee to look into the privacy concern of data. Data in the cloud may be distributed across geographical border and may not be satisfying privacy law of the concerned region, in India Information Technology Act, 2000("IT Act") and its various sub section governs the data security and privacy related issues. Laws prohibit some data from being used for secondary reasons other than the purpose for which it was originally collected.

Data stored on the cloud subject to the legal requirements of one or more regulations—for instance, The Health Insurance Portability and Accountability Act (HIPAA) or The Gramm-Leach-Bliley Act (GLBA) the cloud provider must protects the privacy of the data in the appropriate manner.

**6.3 Data integrity:** Integrity means that assets can be modified only by authorized parties or in authorized ways and refers to data, software and hardware. Data Integrity refers to protecting data from unauthorized deletion, modification or fabrication [3].

In traditional system data integrity can be easily achieved. Data integrity in such a system is maintained via database constraints and transactions. Transactions should follow ACID (atomicity, consistency, isolation and durability) properties to ensure data integrity. ACID properties are supported by most of the database system software like Oracle, SQL and ensure integrity.

Data generated by cloud computing services are kept in the clouds. Keeping data in the clouds means users may lose control of their data and rely on cloud operators to enforce access control and integrity.

**6.4 Data Location:** Location of data is also concerned in the cloud computing. The Cloud Service Providers (CSPs) have their datacenter located across the different geographical region, because of the data privacy law in various countries this may be a matter of concern. Some data may be of sensitive nature and they must not leave certain country. In case of investigation some data may be required and accessing those data may be a problem [4].

**6.5 Data Lock in:** Selecting a Cloud Service Provider (CSP) platform that is built on proprietary formats means that businesses can face a lock-in situation, which will make it much more difficult to change the service provider at some point in the future. Cloud Service Provider (CSP) needs to be changed if the cloud provider changes their terms of service or has interruptions to service that cause you to search for alternatives.

Lock-in has been getting increased attention as cloud service providers acquire customers at a rapid pace. In a November 2009 report entitled Cloud Security Risk Assessment, the European Network and Information Security Agency (ENISA) highlighted lock-in as one the biggest risks involved with cloud computing. "There is currently little on offer in the way of tools, procedures or standard data formats or services interfaces that could guarantee data and service portability" said the report. This can make it difficult for the customer to migrate from one provider to another or back to an in-house environment.

Data Lock-In problem can be overcome with the help of CDMI (Cloud Data Management Interface) developed by the SNIA (Storage Networking Industry Association). CDMI is the first industry-developed open standard for cloud computing. CDMI includes the ability to manage service levels that data receives when it is stored in the cloud as well as a common interoperable data exchange format for securely moving data and its associated data requirements from cloud to cloud [5].

**6.6 Data Availability:** Availability refers to the property of a system being accessible and usable upon demand by an authorized entity. Data Availability is one of the prime concerns for the organizations. When keeping data at remote systems owned by others, data owners may suffer from system failures of the service provider. If the Cloud goes out of operation, data will become unavailable as the data depends on a single service provider.

It is responsibility of cloud service provider to ensure that cloud customer get service round the clock without any interruption. To ensure this changes may be needed both at hardware and software level. A multi-tier architecture needs to be adopted, supported by a load-balanced farm of application instances, running on a variable number of servers. Resiliency to hardware/software failures, as well as to denial of service attacks, needs to be built from the ground up within the application. At the same time, an appropriate action plan for business continuity (BC) and disaster recovery (DR) needs to be considered for any unplanned emergencies.

**6.7 Data Segregation:** Data Segregation refers to the separation of data to ensure that each cloud customer accesses his information only without affecting other customer's information. Make sure that your Cloud Service Provider (CSP) uses encryption in data segregation or aggregation and test the encryption schemes by security experts.

Cloud provides services to a number of users through Multi-tenancy in which both hardware and software are shared. As a result of multi-tenancy a number of user data can reside at the same location on the same hardware infrastructure. In this environment Intrusion is quite possible. This intrusion can be done either by hacking through the loop holes in the application or by injecting client code into the SaaS system. A client can write a masked code and inject into the application. If the application executes this code without verification, then there is a high potential of intrusion into other's data. The Cloud provider must ensure a clear boundary for each user's data. The boundary must be ensured not only at the physical level but also at the application level. The service should be intelligent enough to segregate the data from different users.

**6.8 Security Policy and compliance:** A security policy should be strong enough to protect people and information, and define the expected behavior by all entities such as all type of users, system administrators in the organization, management, and security personnel. Security manager should be able to monitor, analyze, and investigate organizations' infrastructure. The policy should define and authorize the consequences of violation, define the company consensus baseline stance on security, help reduce risk and help track compliance with regulations and legislation. A number of important factors like inside threats and access controls are needed to be taken into consideration when employing a policy for ensuring security between a cloud hosting provider and a customer [7].

Traditional service providers need Security Certification and audit. To increase customer trust, cloud service providers needs to follow these security audits. There is tremendous pressure on Enterprises to follow a range of regulations and standards such as PCI, HIPAA, and GLBA in addition to auditing practices such as SAS70 and ISO. These security standards are expected to comply by all organization for all type of server whether it is on premise or remotely located physical or virtual sever.

**6.9 Multitenancy:** Multitenancy is a very important characteristics of cloud computing. Multiple users can run software on the same hardware resources, its increases the resource utilization, despite its benefits it poses security and privacy threats. Using Virtualization technique Cloud customers are isolated from each other but a customer can access to other customer's actual or residual data, network traffics, and operations by exploiting bugs in the applications. Attackers may target shared resources like hard disk, memory, CPU caches, for resource management in the cloud, different virtualized application instances must be constantly provisioned, allocated, or even migrated between multiple physical computer located off the premises or on premises. Due to this attribute of cloud computing (dynamic provisioning) task of maintaining security also increase. Cloud Security Alliance suggested [8] remedies to mitigate this threat such as: conducting vulnerability scanning and remediation, promoting strong authentication and monitoring unauthorized activities and implementing security best practice for installation and configuration. Multitenancy security and privacy is one of the critical challenges for the public cloud, and finding solutions is pivotal if the cloud is to be widely adopted.

**6.10 Non Repudiation:** This refers requirements of preventing a party to an interaction with the cloud to deny the interaction. Digital signature may be a tool to fix responsibilities of an interaction. In [9] the researcher uses a technique to locate the user and get their origin. It enables the storing of users information and makes it very difficult for the users to deceive about their identity information. Multi-party non repudiation (MPNR) protocol [10] provides a fair non repudiation storage cloud and also prevents roll-back attacks.

**6.11 Energy Management:** Cloud infrastructure should be energy efficient and environment friendly. It is found that Powering and cooling of data centre alone accounted for 53 % of total operational cost [11]. Lowering down energy consumption not only save cost but also will be environment friendly. Cloud Service

Providers (CSPs) are now paying attention for energy efficient data center. They are working in several directions. For example, energy efficient hardware architecture that enables slowing down CPU speeds and turning off partial hardware components [12] has become commonplace. Energy-aware job scheduling and server consolidation [13] are two other ways to reduce power consumption by turning off unused machines. Current research focuses not only lowering energy requirement but also improve application performance.

**6.12 Governance and regulatory Compliance:** Cloud Service Provider (CSP) are not only responsible for providing Infrastructure and their maintenance but also they have to incorporate and follow region specific government's rules and regulation such as SOX, HIPAA, FISMA, FIPS 140-2, GLBA, ITAR, ISAE 3402, SAS 70.

Important Governance & regulatory policies are

- SOX Sarbanes-Oxley Act
- HIPAA The Health Insurance Portability and Accountability Act (HIPAA) of 1996
- FISMA The Federal Information Security Management Act of 2002
- FIPS 140-2 The Federal Information Processing Standard (FIPS) Publication 140-2
- GLBA The Gramm-Leach-Bliley Act
- ITAR International Traffic in Arms Regulation
- ISAE3402 INTERNATIONAL STANDARD ON ASSURANCE ENGAGEMENTS (ISAE) 3402
- SAS70 Statement on Auditing Standards

**6.13 Insecure API:** Cloud computing providers expose a set of software interfaces or APIs that is detailed structural documents that customers use to manage and interact with cloud services. All the operation of the cloud like Provisioning, management, orchestration, and monitoring are all performed using these interfaces. These API as publically available intruders may use weakness in these API to compromise security of Cloud Infrastructure.

API Specification are published by cloud provider to the general public because of

1. They want to tell the services those are available.

2. To facilitate customer to redesign there architecture to use services efficiently.

API is publically available so they must be carefully designed so that there should be no bug and legitimate user can use cloud services without any difficulty. They should take proper care in deciding up to what level or extents user should be aware with the functionalities.

**6.14 Service level agreement:** Service level agreement (SLA) is a document which talks about the expected level of services provided by the cloud provider to the cloud User. It defines qualities, priorities and responsibilities. It typically also sets out the remedial action that will be taken if service fall below the level as stated in SLA. Cloud user may setup a committee to work out SLA. It should satisfy users requirement in cost effective manner Cloud users should define their needs clearly before signing the contracts.

This is clearly an extremely important legal contract between a cloud service provider and the cloud user that should have the following qualities [14]

- Identify and define services required by the user.
- Identify security and privacy requirements.
- Identify priorities of the services.
- Identify responsibilities of cloud user and providers
- Reduce areas of conflict
- Unambiguous definition of expectations and obligations on both sides.
- Identify legal and regulatory compliance.

**6.15 Trustworthy Service Metering:** Cloud Service Providers uses different parameters to charges the consumers on the basis of the amount of services they have consumed. For example, the Amazon Elastic Compute Cloud (EC2) charges users based on the time that their specified EC2 instances are in a running state, while Google AppEngine charges on the basis of how many CPU cycles a user application consumes. However, because users might have little or no visibility into the cloud infrastructure, they're often unable to directly connect their actual cloud resource consumption and the usage charges, In cloud computing multiple users share different resources those are not perfectly isolated and software bug or intruders may use the services and there uses charges has to be paid by users. So, how to guarantee service muttering's trustworthiness is very important if the cloud computing paradigm is to be successful.

# [VII] LIMITATIONS / WEAKNESS OF CLOUD COMPUTING

Besides above issue cloud computing has other limitations too such as downtime, Data transfer limitation, support response time, heightened latency, Limited control, Lack of knowledge and difficulty in integrating devices.

**7.1 Outage:** Outage and downtime is possible even to the best cloud service providers, even internet connection at user end may create problem.

**7.2 Data Transfer Limitation:** To transfer large amount of data may be a problem.

**7.3 Support Response Time:** If response to User questions are slow than user may face problem.

**7.4 latency:** Amount of time to takes for your computer to interact with the servers is known as latency. If interaction is very slow than this may be a problem.

**7.5 Limited control:** Since the services run on remote server users have limited control over the function and execution of the hardware and software even cloud software may provide less functionality than locally available software.

**7.6 Lack of Understanding:** Limited accessibility about working of cloud server may be a problem for user.

**7.7 Integration:** Integrating equipment such as printers, mobile devices, and portable storage units can be a problem for user.

## [VIII] CONCLUSION AND FUTURE WORK

In the cloud computing paradigm the application and data are moved to the remote location, run on the virtual computing resource with the help of virtual machine. This unique characteristic, however, poses many security privacy and trust challenges.

In this paper we have reviewed a number of research papers particularly those dealt with security issues and found that security is the main concern that is delaying cloud computing adoption.

Future work on the cloud computing is to proposed a security framework that should address the issues like Data Security, Data Privacy, Data Integrity, Trust management, multi Tenancy, Non repudiation, Secure API, Data Lock-in, Energy Management, SLA, Trustworthy service metering.

## REFERENCES

[1] S. Subashini n, V.Kavitha [2011]. A survey on security issues in service delivery models of cloud computing. Journals of network and computer

Applications. Volume 34, Issue 1, January 2011. Pp.1-11.

[2] Dawei Suna, Guiran Changb, Lina Suna and Xingwei Wanga [2011]. Surveying and Analyzing Security, Privacy and Trust Issues in Cloud Computing Environments. ELSEVIER . Pp. 2852-2856.

[3] Dimitrios Zissis, Dimitrios Lekkas [2012]. Addressing cloud computing security issues. Future Generation Computer Systems, Volume 28, Issue 3 . Pp. 583-592.

[4] Rabi Prasad Padhy1 Manas Ranjan Patra2 Suresh Chandra Satapathy [2011]. Cloud Computing: Security Issues and Research Challenges. IJCSITS Vol. 1, No. 2, December 2011. Pp. 136-146.

[5] http://www.snia.org.au/assets/documents/the danger of cloud lockin cs.pdf

[6] Kui Ren, Cong Wang, and Qian Wang .Security Challenges for the Public Cloud [2012], IEEE INTERNET COMPUTING, January/ February 2012. Pp. 69-73.

[7] Mathisen E. [2011]. Security Challenges and Solutions in Cloud Computing. In Proceedings of the IEEE International Conference on Digital Ecosystems and Technologies. Pp. 208-212

[8] https://downloads.cloudsecurityalliance.org/initiatives/guidance/csaguide.v3.0.pdf.

[9] Z. Shen and Q. Tong [2010]. The security of cloud computing system enabled by trusted computing technology. ICSPS : V2-11–V2-15.

[10] J. Feng, Y. Chen, D. Summerville, W.-S. Ku, and Z. Su [2011]. Enhancing cloud storage security against roll-back attacks with a new fair multiparty non-repudiation protocol, CNNC. Pp. 521–522.

[11] Hamilton J. [2009]. Cooperative expendable micro slice servers (CEMS): low cost, low power servers for Internet-scale services In: Proc of CIDR.

[12] Brooks D et al [2000]. Power-aware micro architecture: design and modeling challenges for the next-generation microprocessors, IEEE Micro. Pp. 26-44

[13] Vasic N et al [2009]. Making cluster applications energy-aware. In: Proc of automated ctrl for data centers and clouds.

[14] http://www.etsi.org/deliver/etsi_tr/103100_103199/103125/01.01.01_60/tr_103125v010101p.pdf.